\documentclass[conference]{IEEEtran}
\IEEEoverridecommandlockouts
\usepackage{subcaption}
\usepackage[pdftex]{graphicx}
\usepackage{cite}
\usepackage{upgreek}
\usepackage{cite}
\usepackage{xcolor,soul,framed} 
\colorlet{shadecolor}{yellow}
\usepackage{url}
\usepackage{array}
\usepackage{caption}
\usepackage{balance}
\colorlet{shadecolor}{yellow}
\usepackage{color,soul}
\usepackage[cmex10]{amsmath}
\usepackage{algorithmic}
\usepackage{mdwmath}
\usepackage{mdwtab}
\usepackage{eqparbox}
\usepackage{array}
\usepackage{fixltx2e}
\usepackage{dblfloatfix}
\usepackage[mathscr]{euscript}
\usepackage{amsfonts}
\usepackage{amsmath}
\usepackage{commath}
\usepackage{upgreek}
\usepackage[long]{optidef}
\usepackage{bm}
\usepackage{algorithm}

\usepackage[open,openlevel=2]{bookmark}
\usepackage{cleveref}
\usepackage{textgreek}
\usepackage{booktabs,chemformula}
\usepackage{float}
\usepackage{amssymb}
\usepackage{bm}
\usepackage[thinlines]{easytable}
\usepackage{comment}
\usepackage{array, makecell}
\newcolumntype{P}[1]{>{\centering\arraybackslash}p{#1}}

\begin{document}
\DeclareGraphicsExtensions{.png}
\IEEEoverridecommandlockouts
\title{HARQ Retransmissions in C-V2X: A BSM Latency Analysis} 
\author{\IEEEauthorblockN{Abdurrahman Fouda$^{\dag}$, Randall~Berry$^{\dag}$
and Ivan~Vukovic$^{\ddag}$ 
\thanks{This project was supported in part by the Ford-Northwestern University Alliance.}
}
\IEEEauthorblockA{$^{\dag}$Department of Electrical and Computer Engineering, Northwestern University, Evanston, IL\\ 
$^{\ddag}$Ford Motor Company, Dearbon, MI\\
abdurrahman.fouda@u.northwestern.edu, rberry@northwestern.edu, ivukovi6@ford.com}}

\markboth{}%
{Shell \MakeLowercase{\textit{et al.}}: Bare Demo of IEEEtran.cls for IEEE Journals}

\maketitle
\vspace{-0.5in}
\begin{abstract}
 Cellular vehicular-to-everything (C-V2X) systems offer the potential 
 for improving road safety, in part through the exchange of periodic basic safety messages (BSMs) between nearby vehicles. The reliability and latency of these messages is a key metric. Hybrid automatic repeat request (HARQ) retransmissions are one technique used to this end. However, HARQ may come at the expense of consuming the limited available wireless resources, especially in highly congested scenarios. This paper studies BSM transmission latency  and reliability  when HARQ retransmissions are used with the semi-persistent scheduling (SPS) in C-V2X transmission mode 4. We do so through extensive system-level simulations that closely follow the SPS process. Furthermore, we provide an analytical model for the tail behavior of the BSM latency distribution with HARQ retransmissions that is a good approximation to the simulation results.  Our study reveals the impact of several deployment settings (e.g., bandwidth configurations and vehicle density).
\end{abstract}

\IEEEpeerreviewmaketitle

\section{Introduction}\label{sec_intro}
\IEEEPARstart{C}{ellular} vehicle-to-everything (C-V2X) was introduced by 3GPP to support communication between vehicular users (VUEs) and between VUEs and infrastructure~\cite{3gpp22886}. The necessary LTE enhancements to support C-V2X were a study item in Rel-15 \cite{3gpp36885} and continued in Rel-16 for new radio (NR) enhancements ~\cite{3gpp38885}. 3GPP adopted two transmission modes (mode 3 and mode 4) to support C-V2X (also known as LTE-V2X) with and without assistance from base stations (eNBs). This paper focuses on mode 4, which uses a decentralized approach for managing radio resources without assistance from eNBs called semi-persistent scheduling (SPS) ~\cite{spsjour}. 
Application layer standards for C-V2X have been defined by SAE, including the definition of basic safety messages (BSM) for periodic updates exchanged between nearby VUEs~\cite{J3161, J2945}. 

C-V2X supports hybrid automatic repeat request (HARQ) retransmission to improve the packet reception ratio (PRR) and and Inter-packet gap (IPG) of BSM transmissions~\cite{3gpp36321}. However, HARQ retransmissions consume more resources, which may impact performance in high VUE density scenerios.  Our goal in this paper is to study the impact of HARQ in terms of PRR and IPG.  

The IPG performance of SPS scheduling has been studied  in~\cite{vnc} and~\cite{qcom_1shot}, where the role of {\it one-shot transmissions} has also been highlighted. The IPG and PRR performance of C-V2X in dense highway scenarios was compared with that of the DSRC (another vehicular communication protocol) in~\cite{qcom_cc}. An alternative method to decrease the probability of losing consecutive BSMs has been discussed in~\cite{bspots}. The performance of a blind retransmission method has been evaluated in~\cite{blindretransmission} for C-V2X with varying traffic speeds and for NR V2X in~\cite{nrv2x}. However, no previous literature has studied the impact of HARQ retransmissions on the IPG distribution in C-V2X as we do here.

This paper considers studying the tail distribution of IPG CCDFs when HARQ retransmissions are used with the sensing-based SPS and one-shot reselections for BSM transmission in C-V2X networks. IPG and PRR performances are evaluated through extensive simulation campaigns using a C++ system-level simulator that closely follows the C-V2X transmission mode 4 process. Our analysis reveals that the best IPG tail improvement is achieved when HARQ retransmissions and one-shot reselections are used at small vehicle-to-vehicle (V2V) distances with high VUE densities and at any V2V distance with moderate and low densities. Also, it is shown that using HARQ retransmissions and one-shot reselections yields the best PRR performance when the system is noise-limited. Furthermore, we propose an analytical model to characterize the the IPG tail when HARQ retransmissions are used. This builds on our prior work in~\cite{spsjour} which did not include HARQ retransmissions.

\section{SPS with one-shot reselection and HARQ}\label{sec_tm4}

We consider a setting where SPS interleaved with one-shot reselections is used to allocate virtual radio resource blocks (VRBs) for BSM transmissions.  VRBs are autonomously (re)selected from a candidate list which is defined using a pre-configured resource pool of time-frequency resources.\footnote{More precisely, we consider a setting where each VRB consists of 10 physical resource blocks (PRBs) in LTE which occupy a 1 msec subframe.} The candidate VRBs that can be selected depend in part on the measured signal strength in those VRBs.
The VRB reselection process is controlled by the SPS and the one-shot reselection counters $C_{\mathrm{s}}$ and $C_{\mathrm{o}}$, which are chosen uniformly at random between $\left[\alpha_{\mathrm{s}},\,\beta_{\mathrm{s}}\right]$ and $\left[\alpha_{\mathrm{o}},\,\beta_{\mathrm{o}}\right]$, respectively. These counters decrease by one for each transmissions and VRBs are only reselected when the counter reaches zero.  Namely, when $C_{\mathrm{s}}$ expires, VRBs are 
reselected with a reselection probability $p_{\mathrm{r}}$  (see~\cite{spsjour} for further discussion). When $C_{\mathrm{o}}$ expires, a VUE reselects a VRB for a one-shot transmission, and then returns to its previous VRB.  
 
C-V2X mode 4 supports up to one redundant retransmission using HARQ to improve the signal-to-interference-plus-noise ratio received (SINR) and decrease packet losses~\cite{3gpp36321}. VRBs for HARQ transmissions are selected from the same candidate list such that the subframe of HARQ VRBs lies within 15 subframes from the subframe of the initially selected set of VRBs. Whenever a vehicle reselects a VRB (due to either $C_{\mathrm{s}}$ or $C_{\mathrm{0}}$ expiring), it also reselects the VRB used for HARQ transmissions.  

Given that BSMs are transmitted in a broadcast scenario, the HARQ retransmission of a BSM is transmitted regardless of whether its first attempt is received. Similar to the first transmission, the first two physical resource blocks (PRBs) of a HARQ retransmission are reserved for the sidelink control information (SCI) and the remaining PRBs (within two contiguous VRBs) are reserved for the data packets. For each two transmitting attempts of the BSM, the receiving VUE implements HARQ combining for the received SCIs and data packets. Note that a BSM is successfully received if the SINR received from either of the two transmission attempts or the combined is above the SINR thresholds shown in Table~\ref{tab_simpara}.       

We use the SAE congestion control specified in~\cite{J2945},~\cite{J3161} to reduce the probability of packet collision in highly congested scenarios. This is done by adjusting the BSM generation rate based on the estimated number of nearby transmitting vehicles. When congestion control is not active, BSMs are generated once every 100 msec, which is also the underlying frame size from which VRBs are selected. When the estimated number of vehicles within 100 meters exceeds 25, the BSM rate linearly reduces until it reaches a minimum value of one BSM every 600 msec.  In the simulations presented, congestion control is not active when the VUE density is 125 VUE/km and is active for higher densities, with the time between BSM having it maximum value of 600 msec when the density is 800 VUE/km. 

\section{Simulation Results}\label{sec_results}

In this section, we study the IPG distribution and PRR performance when HARQ retransmissions are used in C-V2X systems. IPG is measured as the time elapsed between each successive pair of successfully received BSMs between a pair of vehicles. The value of the IPG CCDF at $i$ milliseconds and a V2V distance of $d$ meters represents the fraction of IPG instances between vehicles separated by $d$ meters that exceed $i$ milliseconds. PRR at a V2V distance of $d$ meters quantifies the transmission reliability by calculating the ratio between the total number of BSM transmitted and successfully received between any pairs of VUEs that are separated by $d$ meters.

IPG and PRR are analyzed using different deployment settings, including different VUE densities, one-shot settings, V2V distances, and bandwidth configurations. We use a high-fidelity system-level simulator that closely follows the sensing-based SPS for BSM transmission using C-V2X transmission mode 4. A single-lane highway scenario is used to carry out Monte Carlo simulations. Numerical statistics are collected from vehicles located only in the middle third of the highway to minimize the boundary effects. The simulation settings are summarized in Table~\ref{tab_simpara}.

\begin{table}[b]
\renewcommand{\arraystretch}{1.3}
  \centering
  \caption{Simulation parameters.}\label{tab_simpara}
  \centering\renewcommand\cellalign{lc}
    \setcellgapes{500pt}\makegapedcells
        \begin{tabular}{l l}
        \hline
            \textbf{Parameter}  & \textbf{Value}\\
        \hline
            Scenario & Layout: 5 km single-lane Highway\\ 
            & Density: 125, 400 and 800 VUE/km\\
        \hline
            Channel model  & Path loss: ITU-R UHF urban canyon~\cite{toyota}\\  
            & In-band emission: 3GPP TR 36.885~\cite{3gpp36885}\\
            & Fast fading: Nakagami-$m$ distribution\\
        \hline
            Antenna settings & $n_t=1$, $n_r=2$, MRC receiver\\
        \hline 
           & $C_{\mathrm{s}}\in{[5,\,15]}$, $p_{\mathrm{k}}=0.8$\\
        \hline
            Resource pool settings & Carrier bandwidth: 10, 20 MHz at 5.9 GHz\\
        \hline
            Power settings   & Tx power: 20 dBm, Noise figure: 6 dB\\
            & PSCCH power boost: 3dB \\
        \hline
            SINR threshold & PSSCH: 3 dB, PSCCH: 0 dB\\
        \hline
            HARQ settings  & Maximum retransmissions: 1\\ 
            & HARQ selection window: [-15, 15] ms\\ 
        \hline
            Simulation time & 500 seconds, warm-up: first 10 seconds.\\
        \hline
        \end{tabular}
\end{table}

\begin{figure*}
\subfloat[800 VUE/km]{\includegraphics[width=.32\textwidth]{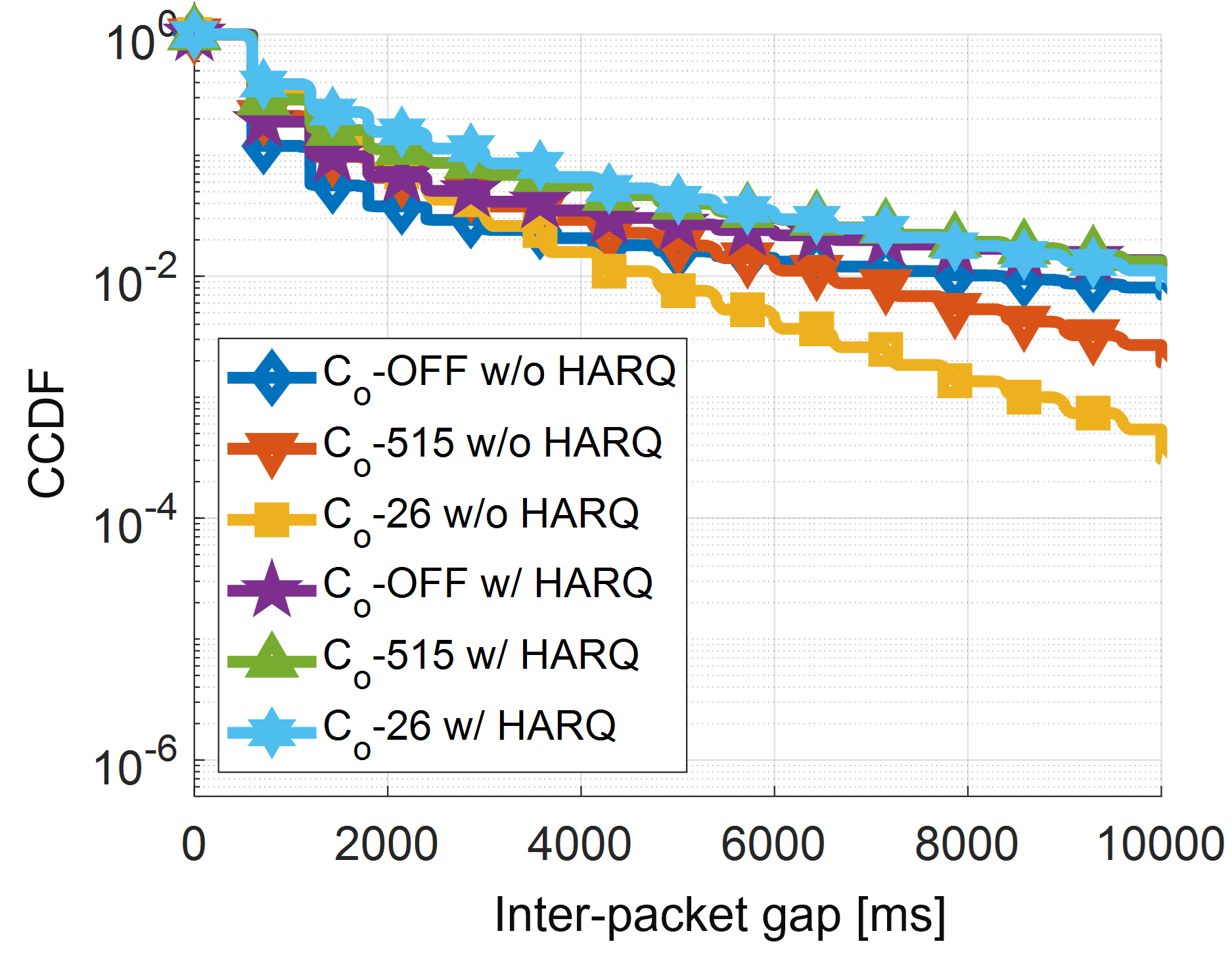}} 
\subfloat[400 VUE/km]{\includegraphics[width=.32\textwidth]{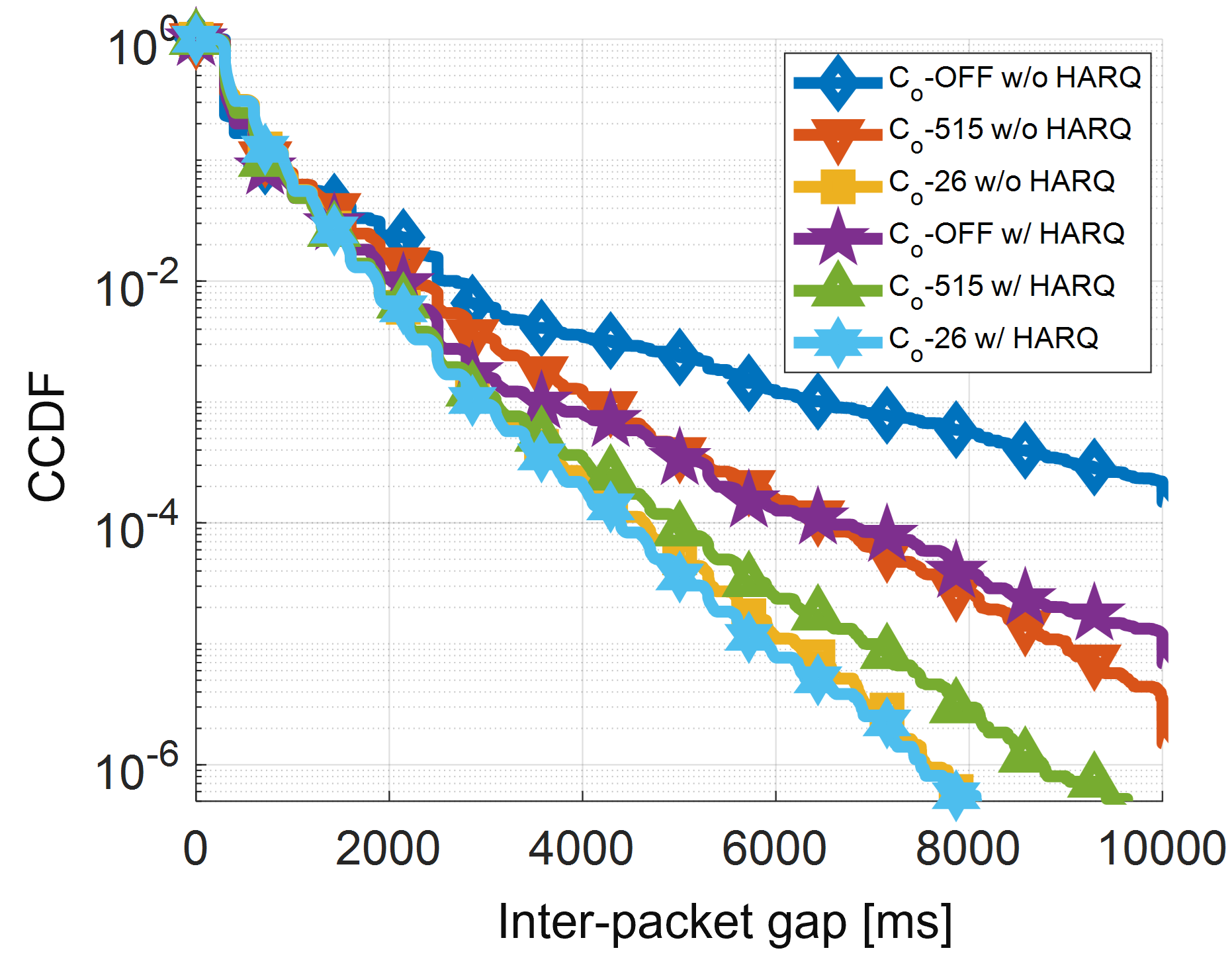}} 
\subfloat[125 VUE/km]{\includegraphics[width=.32\textwidth]{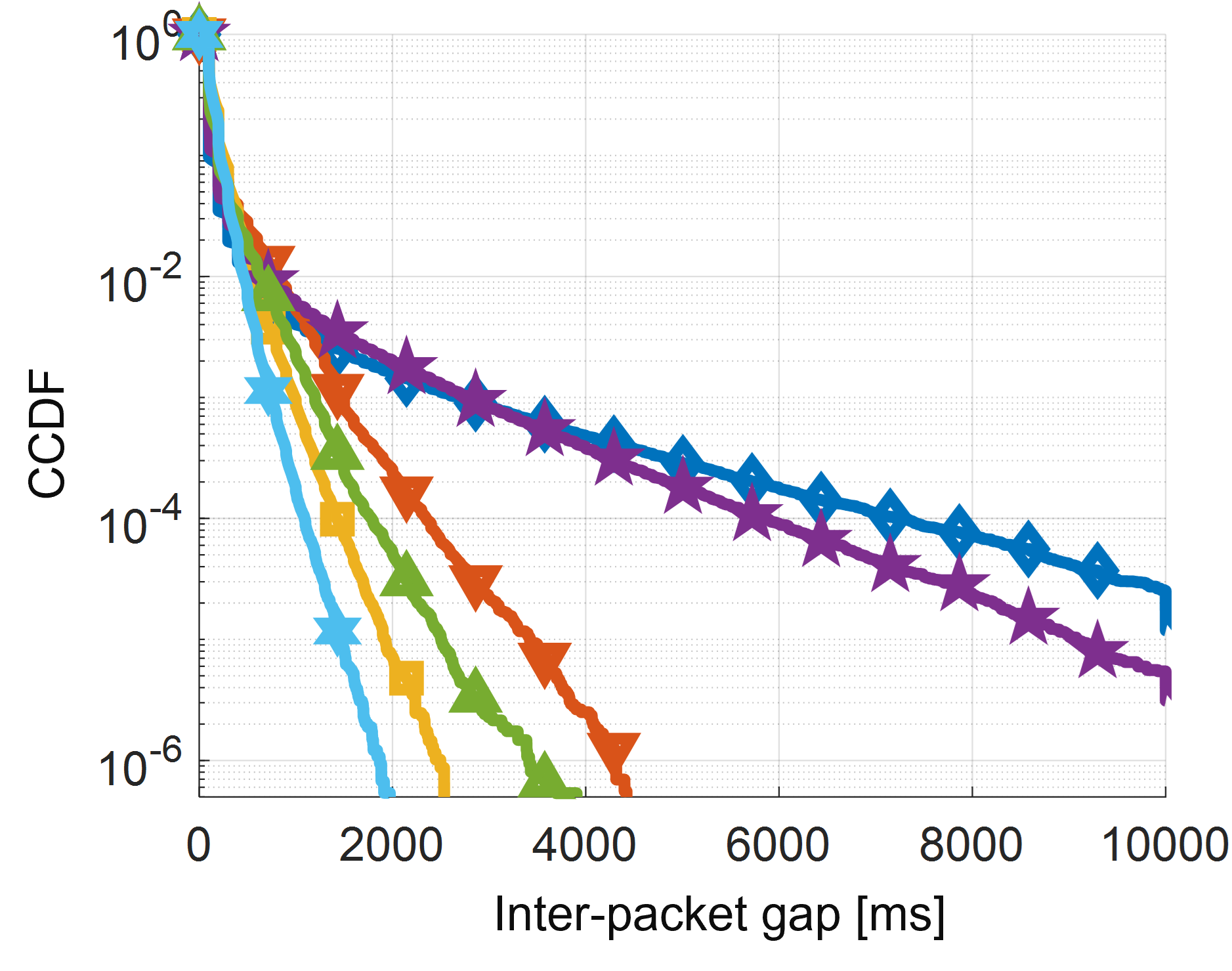}} 
\caption{10 MHz, V2V distance = 300 m, with and without HARQ retransmissions.}
\vspace{-.15in}
\label{fig_10300}
\end{figure*}

\subsection{IPG Performance}\label{sec_sub_ia}

Fig.~\ref{fig_10300} (a) compares the IPG tail distribution of the three configurations of $C_{\mathrm{o}}$, OFF, $[2,\,6]$, and $[5,\,15]$ with and without HARQ retransmissions at a 10 MHz bandwidth for a high VUE density of 800 VUE/km. Figs.~\ref{fig_10300} (b) and (c) show similar comparisons for moderate and low VUE densities of 400 and 125 VUE/km, respectively. As shown in Fig.~\ref{fig_10300} (a), the IPG tails suffer substantial performance degradation when HARQ retransmissions are used. This applies to all three configurations of $C_{\mathrm{o}}$. Specifically, using the one-shot method (either $[2,\,6]$ or $[5,\,15]$ configuration) with the HARQ retransmissions does not improve the IPG tail performance. In fact, deactivating the HARQ retransmissions yields a slightly better tail performance even without using the one-shot method. This is due to the scarcity of VRBs available for BSM transmissions. Essentially, with such a highly congested scenario (800 VUEs/km) and low operating bandwidth (10 MHz), it is expected that the number of available VRBs in the SPS (re)selection will be significantly low. As shown, using the one-shot setting of $[2,\,6]$ without HARQ retransmissions produces the best IPG tail performance.


At lower VUE densities, as shown in Figs.~\ref{fig_10300} (b) and (c),  using the HARQ retransmissions improves the IPG tail distribution in all configurations of $C_{\mathrm{o}}$. Furthermore, the improvement is better with the lower VUE densities (i.e. Fig.~\ref{fig_10300} (c) vs. (b)). The statistics shown in Fig.~\ref{fig_10300} are generated for a V2V distance of 300 m. Similar trends were observed for other V2V distances. 
This suggests that at high VUE densities that HARQ retransmissions may not be productive.

\begin{figure*}
\subfloat[800 VUE/km]{\includegraphics[width=.32\textwidth]{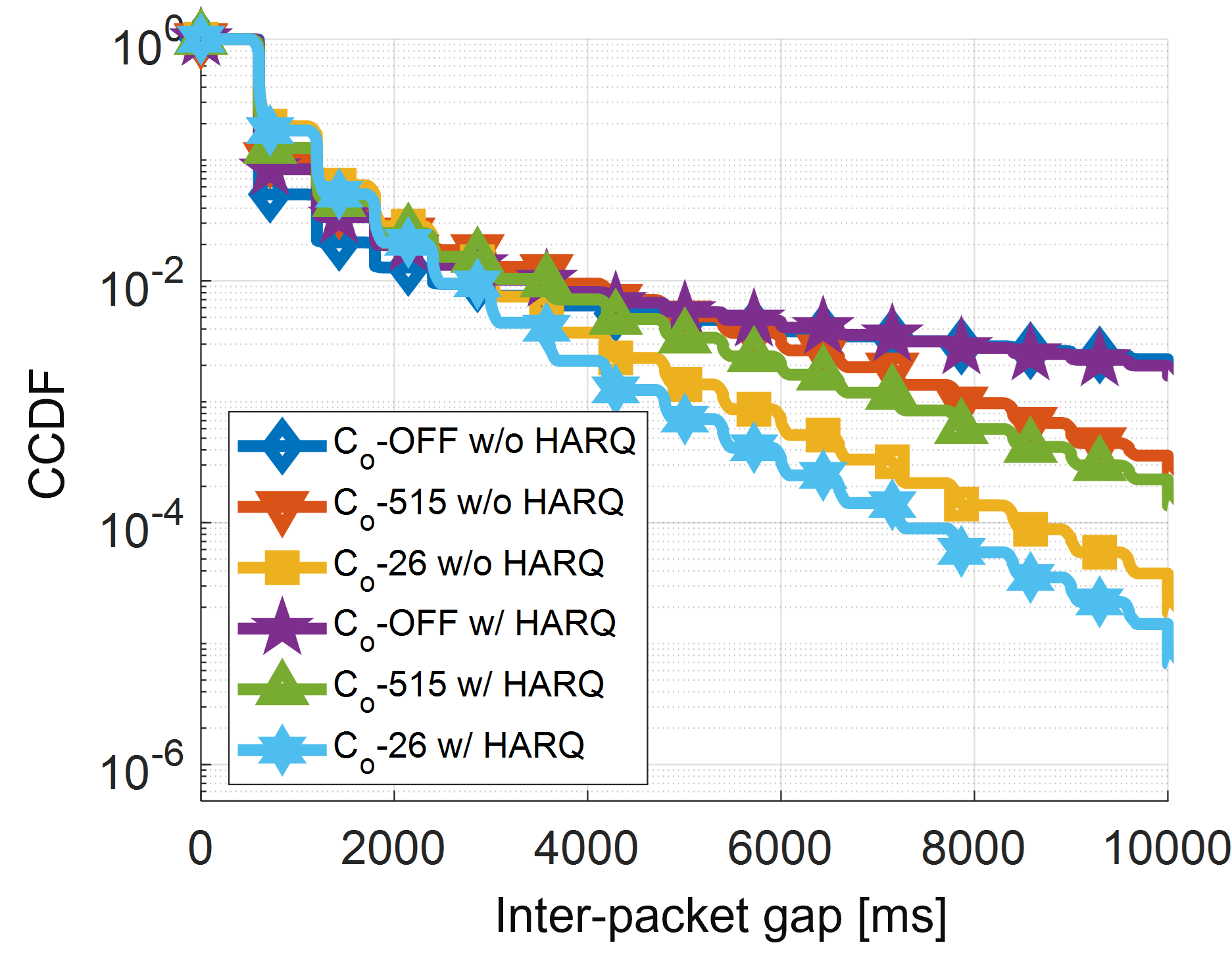}} 
\subfloat[400 VUE/km]{\includegraphics[width=.32\textwidth]{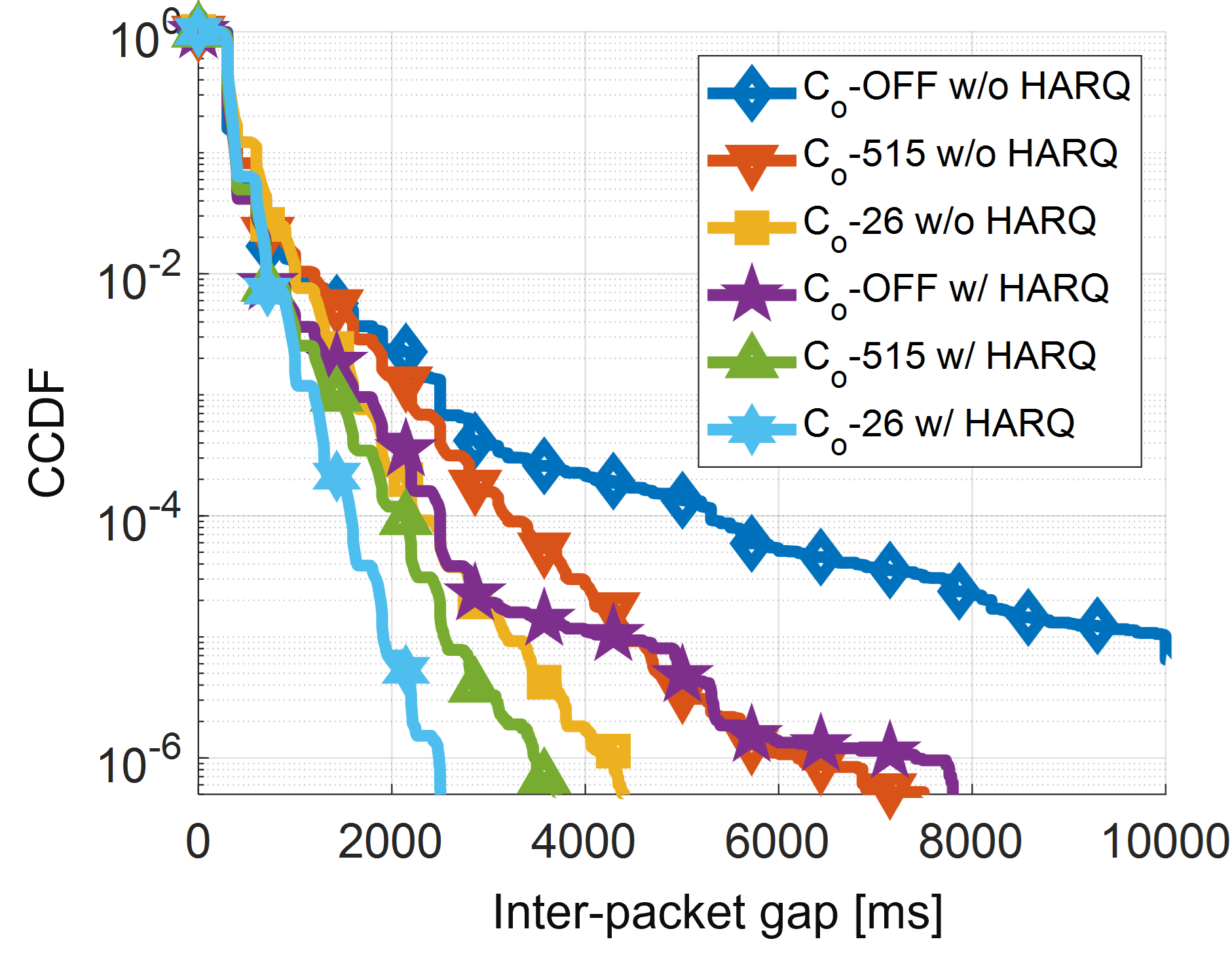}} 
\subfloat[125 VUE/km]{\includegraphics[width=.32\textwidth]{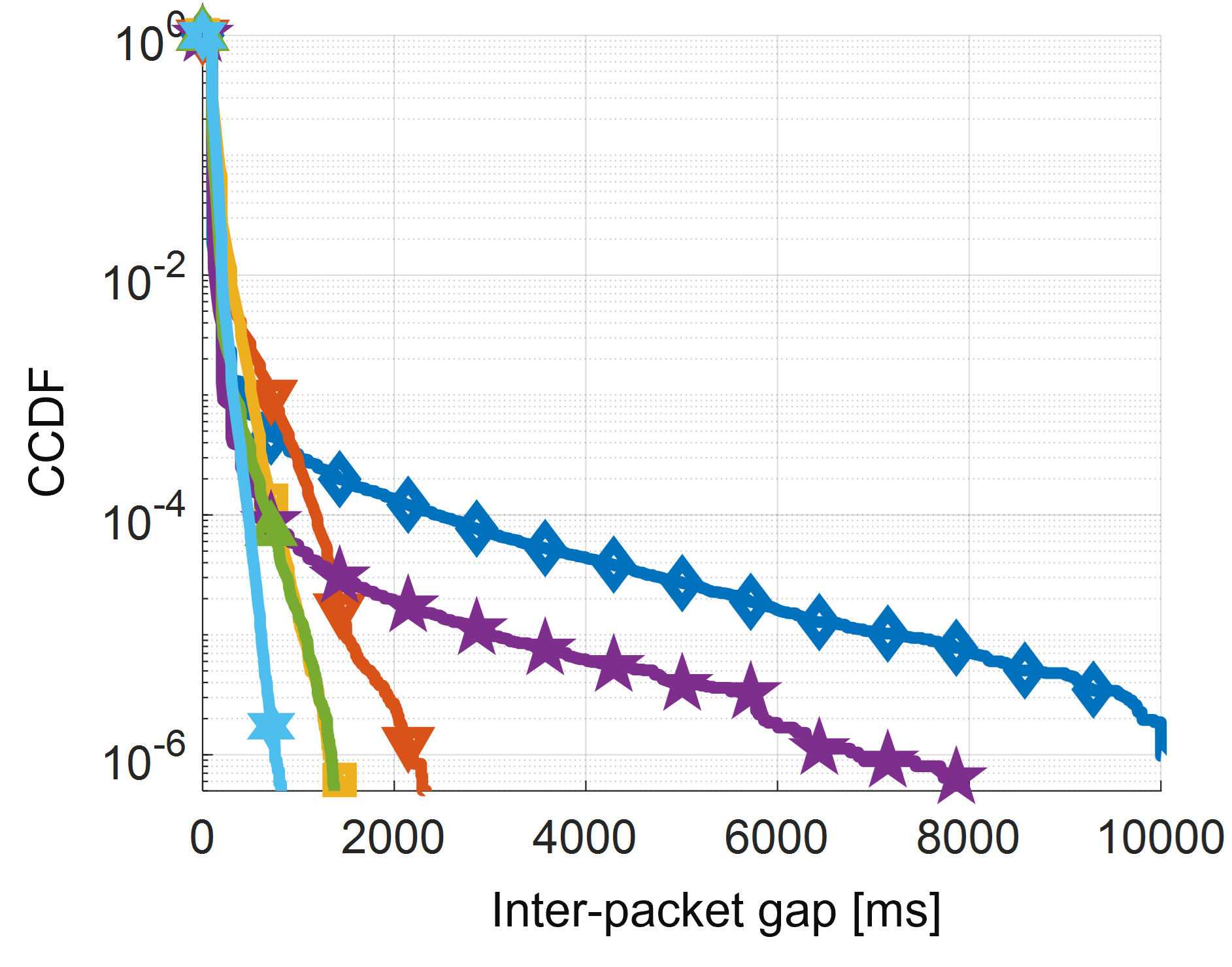}} 
\caption{20 MHz, V2V distance = 200 m, with and without HARQ retransmissions.}
\vspace{-.15in}
\label{fig_20200}
\end{figure*}

However, using HARQ retransmissions with higher bandwidth configurations (e.g., 20 MHz) yields significantly better IPG tail performance, as shown in Fig.~\ref{fig_20200}. It can also be seen that the improvement is greater at lower VUE densities. 
Fig.~\ref{fig_20200} (a) shows that using the one-shot configuration of $[2,\,6]$ without HARQ retransmission significantly improves the IPG tail performance compared to the OFF and $[5,\,15]$ configurations of $C_{\mathrm{o}}$ with HARQ retransmissions. However, using the same configuration (i.e. $[2,\,6]$) with HARQ yields a slightly better tail.
At lower densities (Fig.~\ref{fig_20200} (b) and (c)), the gains are even more apparent.

\begin{figure}
  \begin{center}
  \includegraphics[width=8.85cm,height=8.85cm,,keepaspectratio]{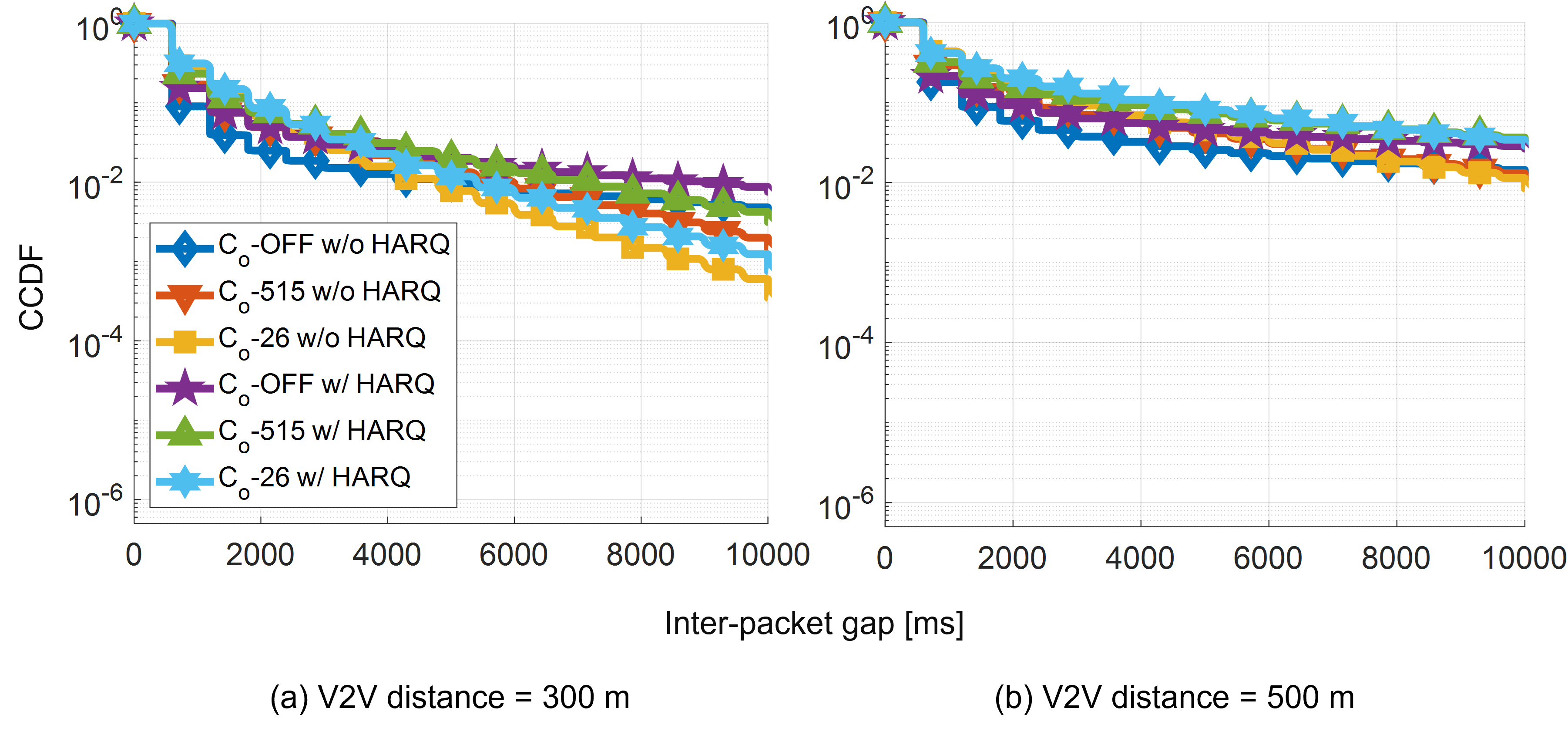}
  \caption{IPG at different V2V densities, 800 VUE/km.}\label{fig_204k300500}
  \vspace{-.15in}
  \end{center}
\end{figure}

Fig.~\ref{fig_204k300500} (a) and (b) depict the effects of using HARQ retransmissions with 800 VUE/km at larger V2V distances of 300 and 500 m, respectively. This shows that even with 20Mhz bandwidth, using the HARQ retransmissions may worsen the IPG tail at larger V2V separation. As shown, the one-shot configuration of $[2,\,6]$ without HARQ retransmissions yields better IPG tail performance than any other configuration of the $C_{\mathrm{o}}$ counter when HARQ retransmissions are used. This shows that the advantages of HARQ depend on both the distance, the bandwidth configuration and the underlying vehicle density.

\subsection{PRR Performance}\label{sec_sub_ia}

\begin{figure*}
\subfloat[800 VUE/km]{\includegraphics[width=.32\textwidth]{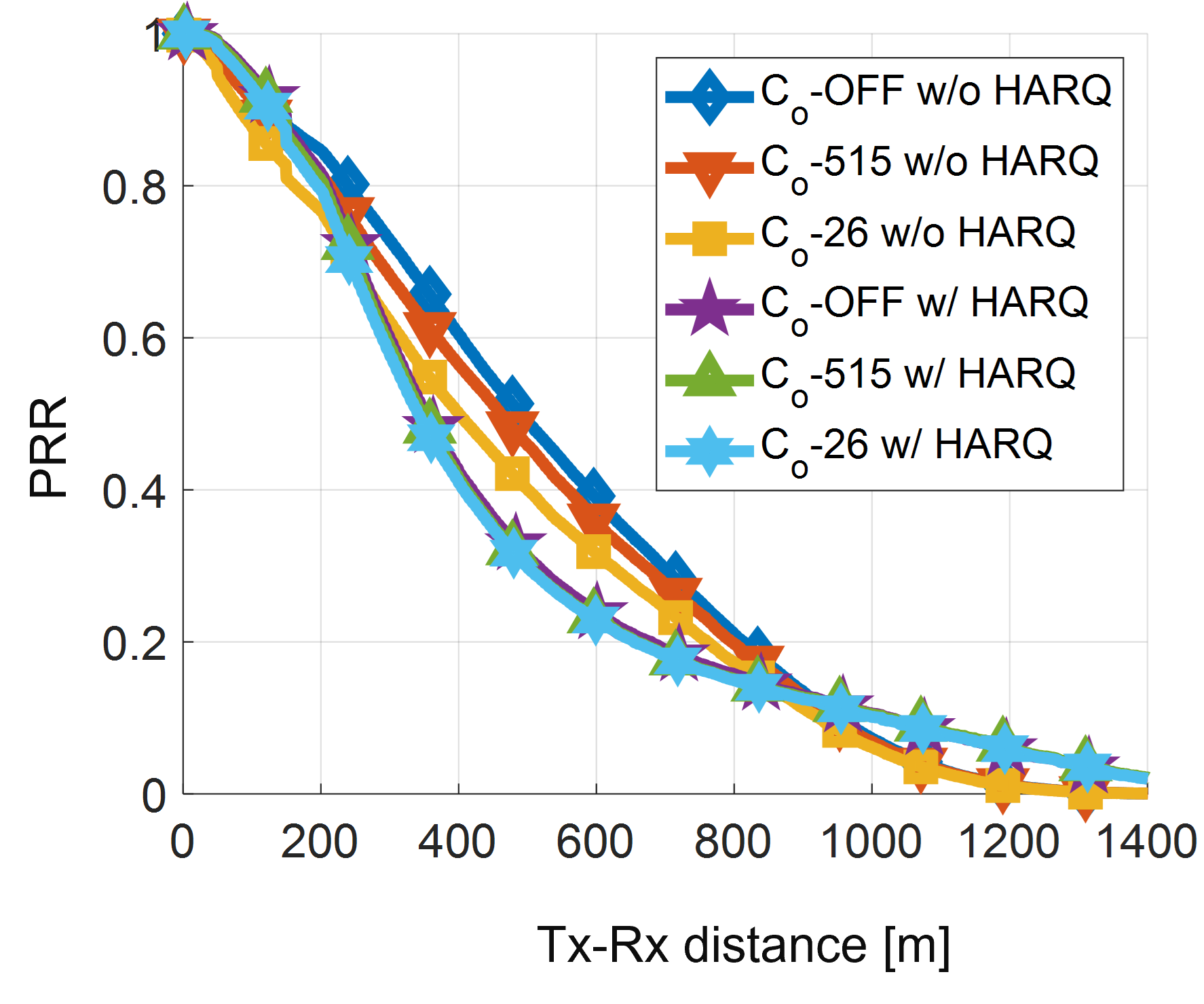}} 
\subfloat[400 VUE/km]{\includegraphics[width=.32\textwidth]{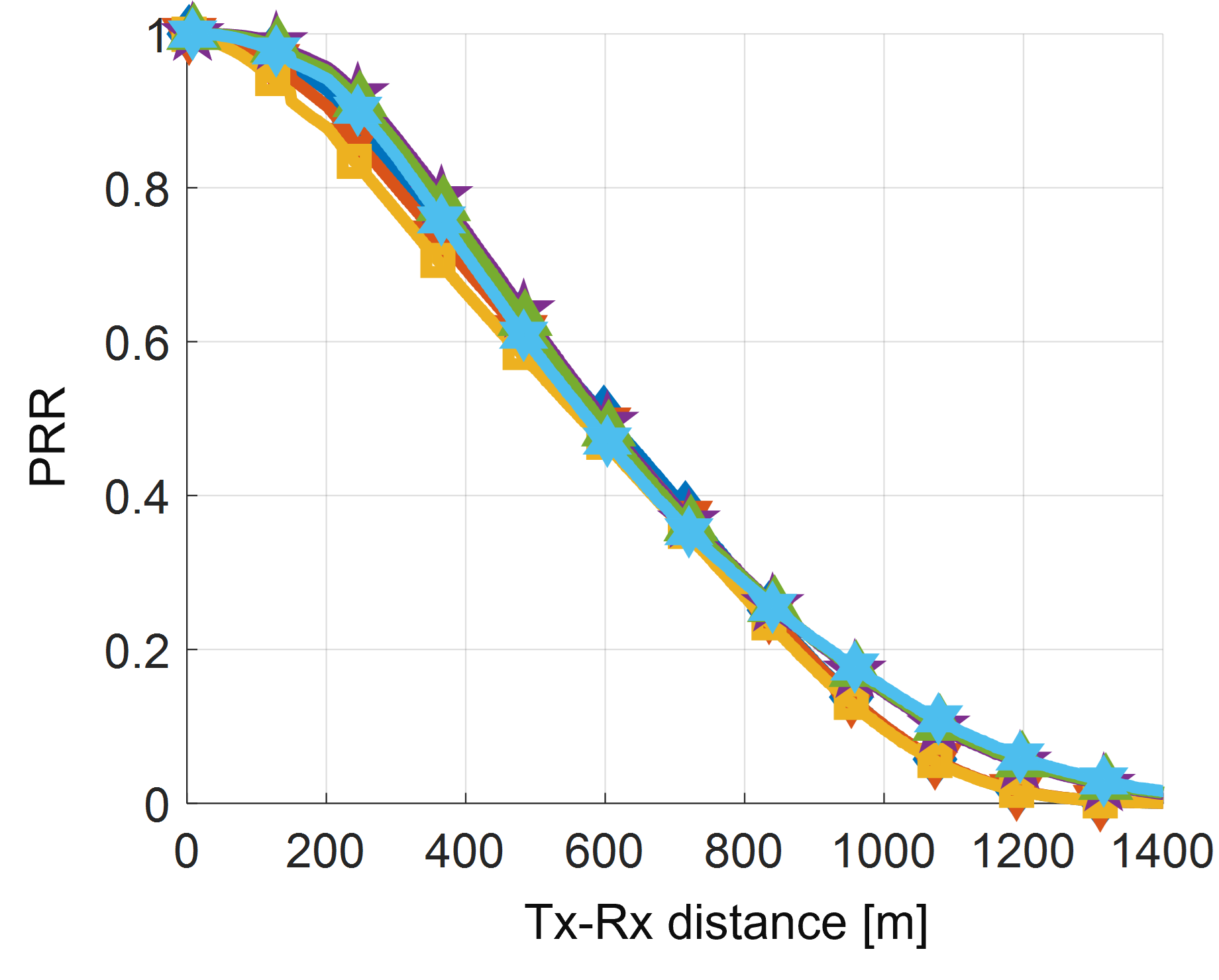}} 
\subfloat[125 VUE/km]{\includegraphics[width=.32\textwidth]{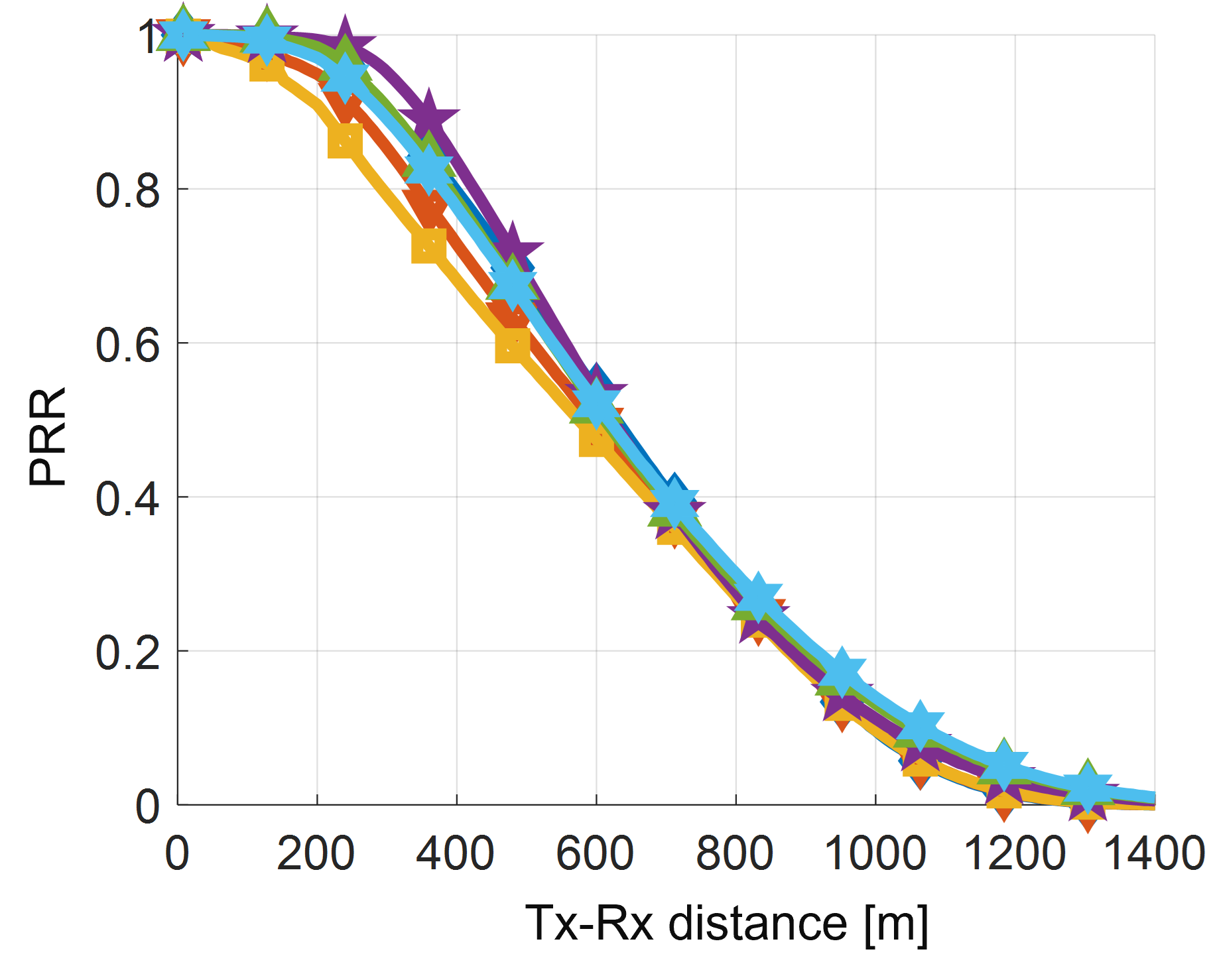}} 
\caption{PRR performance at 20 MHz.}
\vspace{-.15in}
\label{fig_20prr}
\end{figure*}

The PRR performance versus distance of different configurations of $C_{\mathrm{o}}$ and HARQ using a bandwidth of 20 MHz is 
shown in Fig.~\ref{fig_20prr}. As shown, HARQ improves the PRR at large Tx-Rx separations. The improvement is more pronounced at higher VUE densities. This happens because the system is noise-limited at high Tx-Rx separations. On the other hand, Fig.~\ref{fig_20prr} (a) reveals that using HARQ retransmissions degrades the PRR performance at moderate Tx-Rx separations, where the system is interference-limited. 

Essentially, BSM transmissions suffer from a higher probability of packet collision at high VUE densities because of the lack of available VRB resources. Degradation is less pronounced at moderate and light VUE density in Figs.~\ref{fig_20prr} (b) and (c), respectively. It is worth noting that the use of HARQ retransmissions in fact improved the PRR performance at small Tx-Rx separations. In other words, the best PRR performance gains from HARQ are with large Tx-Rx separations at fairly high VUE densities and with small Tx-Rx separations at low VUE densities. 

\begin{figure*}
\subfloat[800 VUE/km]{\includegraphics[width=.32\textwidth]{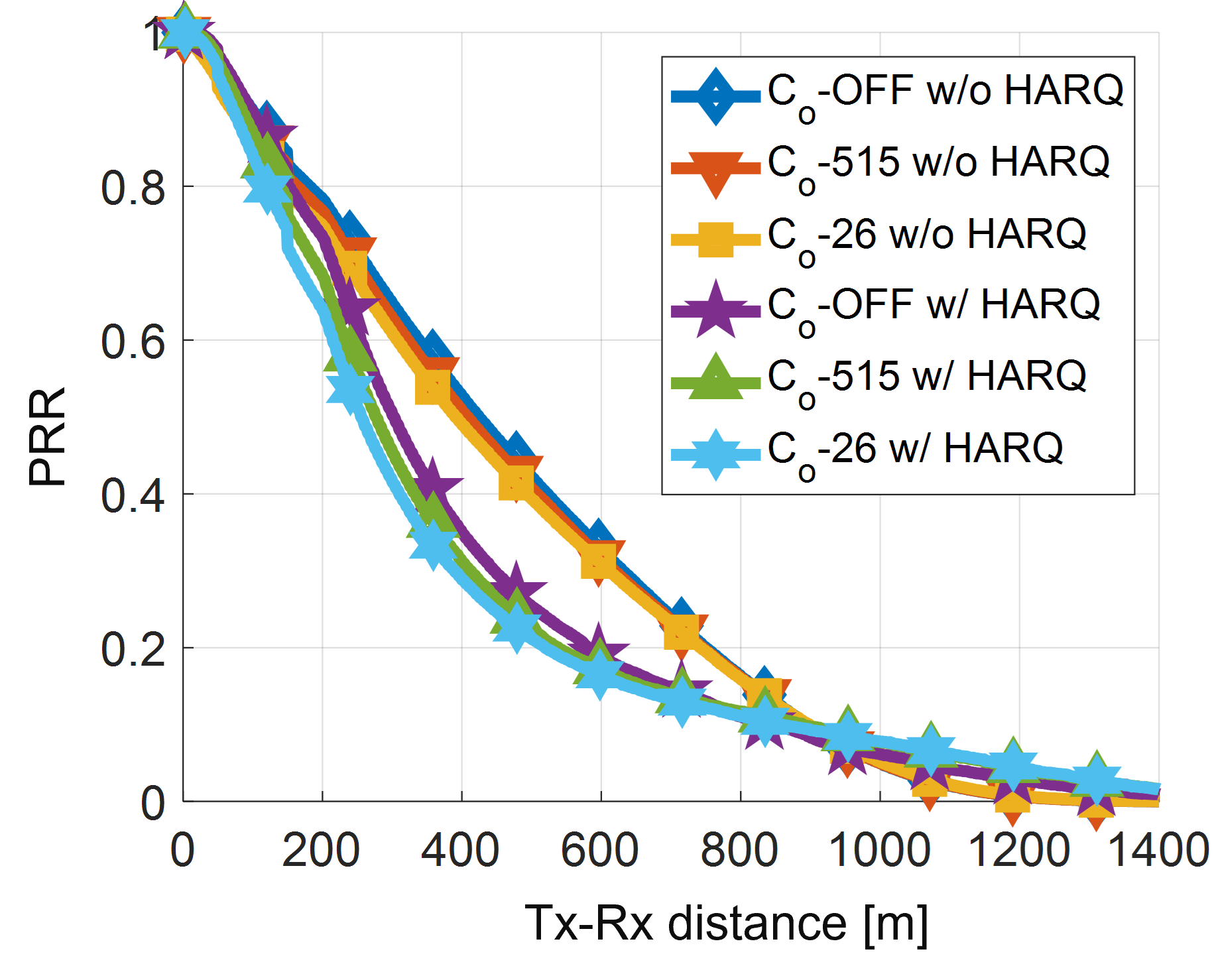}} 
\subfloat[400 VUE/km]{\includegraphics[width=.32\textwidth]{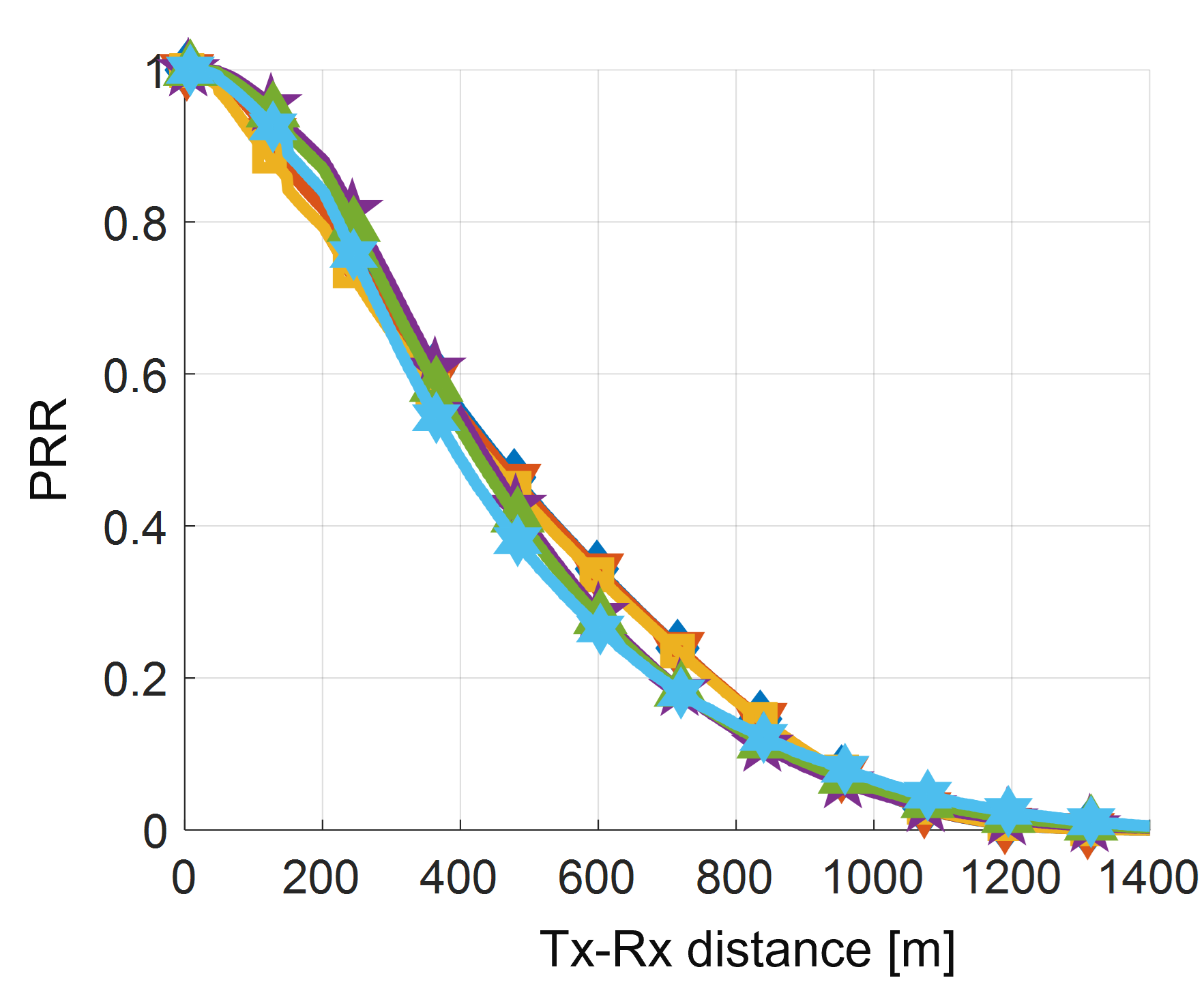}} 
\subfloat[125 VUE/km]{\includegraphics[width=.32\textwidth]{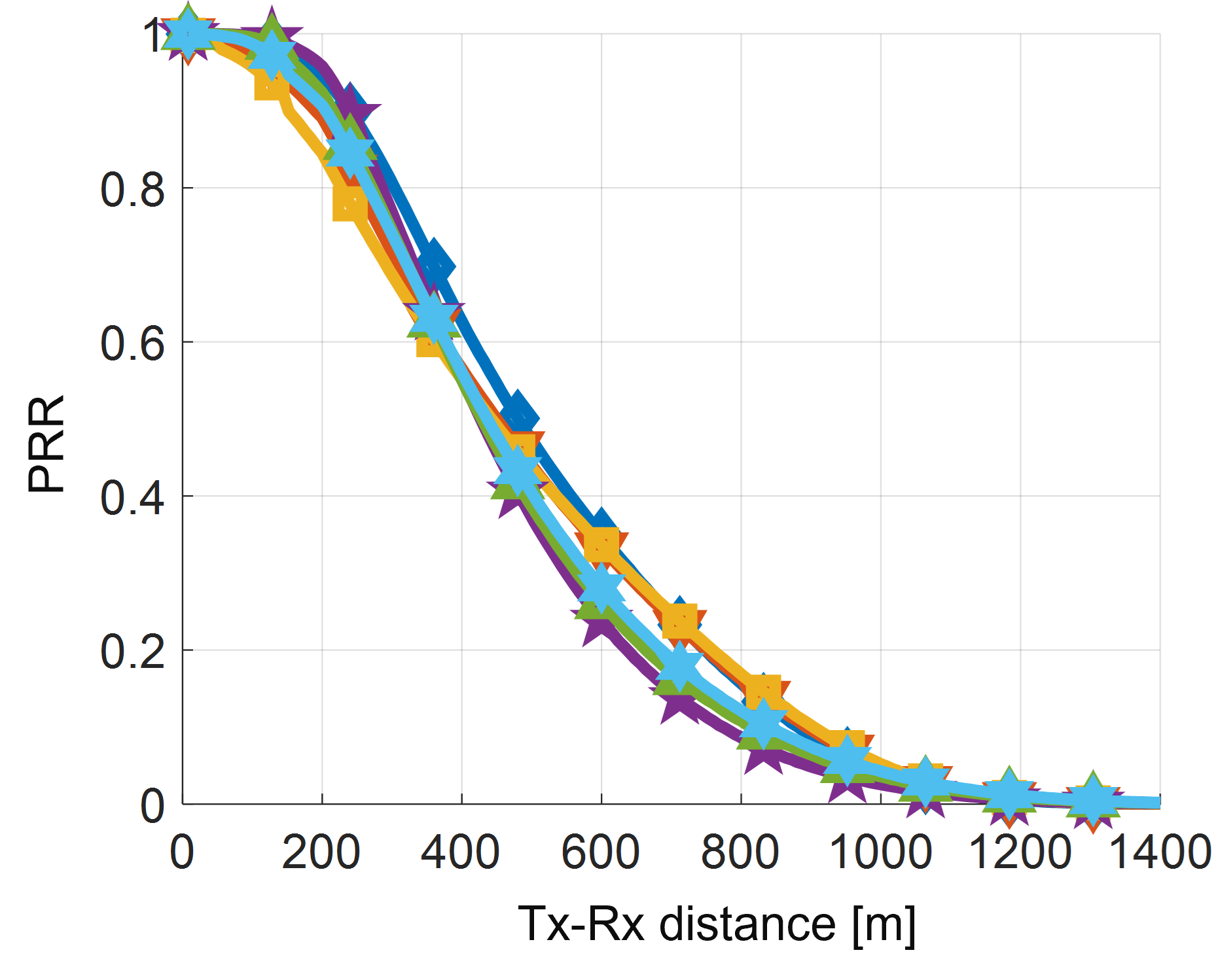}} 
\caption{PRR performance at 10 MHz.}
\vspace{-.15in}
\label{fig_10prr}
\end{figure*}

Similar trends were observed in a 10-MHz bandwidth configuration as shown in Fig.~\ref{fig_10prr}. Figs.~\ref{fig_10prr} (b) and (c) reveal that the PRR performance degrades at moderate Tx-Rx separations with moderate and low VUE densities due to the limited number of available VRB resources compared to the 20 MHz case. Fig.~\ref{fig_10prr} (a) also shows that using HARQ retransmissions with the 10 MHz bandwidth at high VUE densities barely improves the PRR performance at large Tx-Rx separations. This is because, in the 10MHz case, the system is interference-limited at small and large Tx-Rx separations. 

\section{Analytical Model for IPG Tail Distribution}\label{sec_analmodel}
In this section, we present an analytical model to approximate the IPG tail distribution with HARQ for C-V2X communications. 
Our approach is based on the model in~\cite{spsjour} which does not include HARQ. 

Unless mentioned otherwise, a BSM transmission means two BSM transmission attempts representing the regular transmission and the HARQ retransmission. For a given transmitter-receiver pair, conditioned on the fact that the first BSM transmission is not received, let $T$ be a random variable that indicates the length of IPG in terms of the required number of BSM transmission opportunities until a successful transmission. The failure of a BSM transmission means that the receiver did not receive the regular transmission of the BSM and its HARQ retransmission. 

Here, we assume that if a BSM is not received, it is due to interference from a first interferer that is using overlapping VRB(s) with the VRB(s) of the first transmission attempt and from a second interferer that is using overlapping VRB(s) with the VRB(s) of the HARQ retransmission. In other words, interference caused by two dominant interferer (one for each transmission) is the main reason for packet losses. Other reasons for packet losses (e.g., half-duplex transmissions) can be fairly ignored on the basis of numerical data collected from our simulations. In this case, under the SPS, the interference will persist until either one of the interferers or the transmitter reselects another resource or does a one-shot transmission. 

We give a non-uniform geometric model for the SPS and the one-shot reselection processes, in which, the probability of a resource reselection during the $k^{\mathrm{th}}$ BSM transmission opportunity is given by $q^{(k)}(\rho,\sigma,p)$, where $q_\mathrm{s}^{(k)}=q^{(k)}\left(\alpha_\mathrm{s},\,\beta_\mathrm{s},\,p_{\mathrm{r}}\right)$ and $q_\mathrm{o}^{(k)}=q^{(k)}\left(\alpha_\mathrm{o},\,\beta_\mathrm{o},\,1\right)$ for the SPS-triggered and one-shot-triggered reselection, respectively. Given that there cannot be reselections in the first $\rho$ BSM transmission opportunities, we set $q^{(k)} = 0$ for $k \leq \rho$, and for $k > \rho$, we set $q^{(k)}$ so that for large $k$, the expected number of reselections after $k$ opportunities is equal to $k/E(T_r)$ where $E(T_r)$ denotes the expected time between reselections as: 
\begin{equation}
E(T_r) = \left(\rho + \frac{\sigma - \rho}{2}\right) \frac{1}{p_s}.
\end{equation}
In this case, $q^{(k)}$ is selected to satisfy $q^{(k)} (k-\rho) = \frac{2kp_s}{\sigma+ \rho}$. Hence, $q^{(k)} = \left(2kp_s\right)/\left((\sigma+ \rho)(k-\rho)\right)$.
The reselection probability of the SPS interleaved with one-shot transmissions is then: 
\begin{equation}\label{eq_qe}
    q_{\mathrm{\mathrm{i}}}^{(k)}=q_{\mathrm{o}}^{(k)}\left(1-q_{\mathrm{s}}^{(k)}\right)+q_{\mathrm{s}}^{(k)}\left(1-q_{\mathrm{o}}^{(k)}\right)+q_{\mathrm{s}}^{(k)}q_{\mathrm{o}}^{(k)},
\end{equation}
assuming that the two processes are independent.

Next, the probability of a successful BSM transmission (due to a reselection) after $k$ transmission opportunities, where each $k^{\mathrm{th}}$ failure represents the failure of the regular BSM transmission and its HARQ retransmission, is calculated as:
\begin{equation}\label{eq_pu}    p_\mathrm{u}^{(k)}=P_{\mathrm{a}}P_{\mathrm{d}}+P_{\mathrm{n}}P_{\mathrm{r}}+P_{\mathrm{a}}P_{\mathrm{r}}.
\end{equation}
Here, $P_{\mathrm{a}}$ denotes the probability that at least one of the two interferers reselects, $P_{\mathrm{d}}$ is the probability that the transmitter does not reselect, $P_{\mathrm{n}}$ indicates the probability that no interferer reselects new VRBs, and $P_{\mathrm{r}}$ is the probability that a resource reselection is triggered at the transmitter. 

In that, $P_{\mathrm{a}}$ can be calculated as:
\begin{equation}\label{eq_pa}
    P_{\mathrm{a}}=1-\left(1-q_{\mathrm{i}}^{(k)}\right)^{2},
\end{equation}
where $\left(1-q_{\mathrm{i}}^{(k)}\right)$ represents the probability that a single interferer does not reselect new VRBs when the reselection counters expire. Therefore, $\left(1-q_{\mathrm{i}}^{(k)}\right)^{2}$ calculates the probability that neither of the two dominant interferers reselects new VRBs. Similarly, $P_{\mathrm{d}}$ is calculated as $\left(1-q_{\mathrm{i}}^{(k)}\right)$ and $P_{\mathrm{r}}$ as $\left(q_{\mathrm{i}}^{(k)}P_{\mathrm{f}}\right)$, where $P_{\mathrm{f}}$ represents the probability that the transmitter's reselection results in a successful transmission (i.e. it does not reselect into the VRB(s) as another strong interferer). Note that the probability of successful transmission due to one-shot reselection is the same as the average success probability of any transmission (see~\cite{spsjour} for further discussion). Finally, $P_{\mathrm{n}}$ can be given by $P_{\mathrm{n}}=\left(1-q_{\mathrm{i}}^{(k)}\right)^{2}$.

The probability that the IPG exceeds $k$ BSM transmissions (including their HARQ attempts) is then written as:
\begin{equation}\label{eq_tk}
    P(T>k)=\prod^{k}_{i=1}\left(1-p_{\mathrm{u}}^{(i)}\right).
\end{equation}
The analytical model proposed for the tail distribution of the IPG CCDF can then be validated by comparing its approximate slope (which is calculated using~(\ref{eq_tk})) with that obtained through simulations. Here, simulations are carried out for $\approx\mathrm{10^8}$ samples to obtain numerical results with a good level of confidence. Since $T$ is defined as conditioned on one BSM not arriving, there will be a constant offset between $P(T>t)$ and the probability that the IPG exceeds the same amount, which corresponds to the probability that the first BSM does not arrive. Hence, ~(\ref{eq_tk}) captures the slope of the IPG CCDF but not the actual value.

\begin{figure}
  \begin{center}
  \includegraphics[width=6.5cm,height=6.5cm,,keepaspectratio]{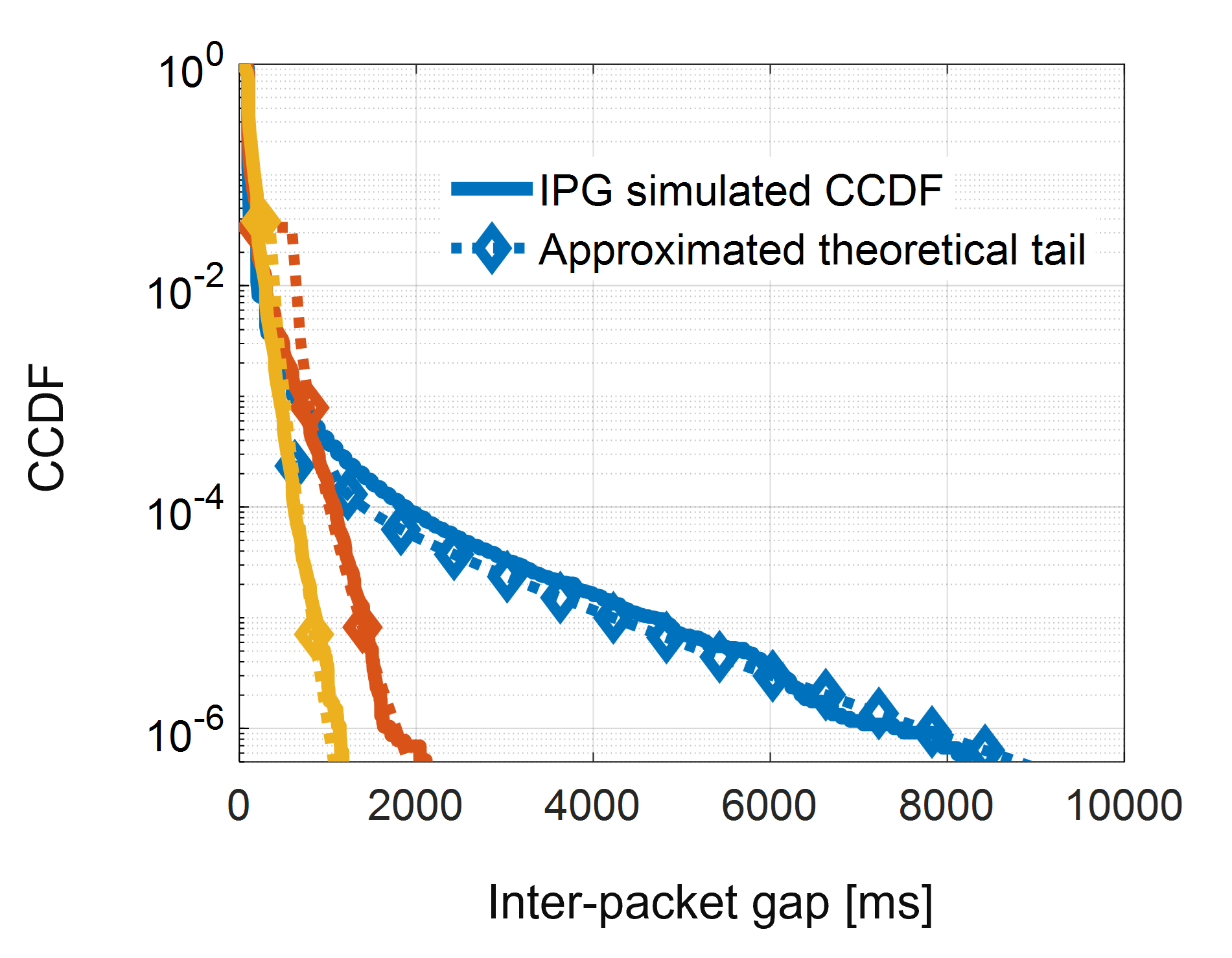}
  \caption{125 VUE/km, 10 MHz.}\label{fig_valid1}
  \vspace{-.15in}
  \end{center}
\end{figure}

\begin{figure}
  \begin{center}
  \includegraphics[width=6.5cm,height=6.5cm,,keepaspectratio]{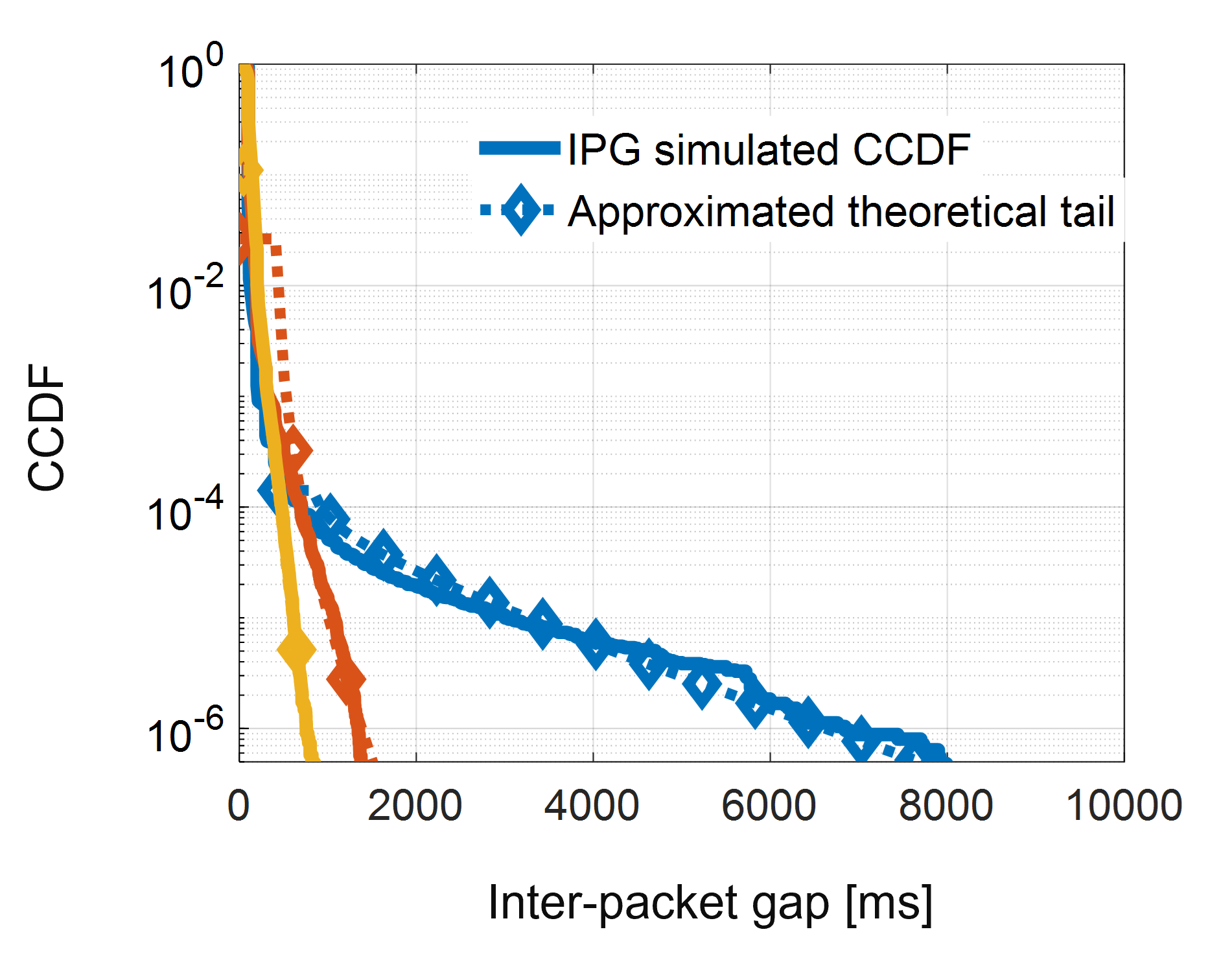}
  \caption{125 VUE/km, 20 MHz.}\label{fig_valid2}
  \vspace{-.15in}
  \end{center}
\end{figure}

Figs.~\ref{fig_valid1} and~\ref{fig_valid2} show the validation of the analytical model for 125 VUE/km using 10 MHz and 20 MHz bandwidths, respectively. Validation is done for the three configurations of $C_{\mathrm{o}}$, OFF, $[2,\,6]$, and $[5,\,15]$. The dotted and solid lines represent the slopes of the analytical and simulation results, respectively. Note that $P_{\mathrm{f}}$ is calculated using Fig.~\ref{fig_20prr} based on the operating scenario and the desired V2V distance bin. Unless otherwise mentioned, the validation presented in this section focuses on the V2V separation of 200 m. Similar trends were observed at other V2V distances. As shown, the IPG tail derived using~(\ref{eq_tk}) gives a reasonable approximation for the actual (i.e., simulated) IPG slopes of different reselection modes at fairly large IPG values (which is the main purpose of the proposed analytical model).

\begin{figure}
  \begin{center}
  \includegraphics[width=6.5cm,height=6.5cm,,keepaspectratio]{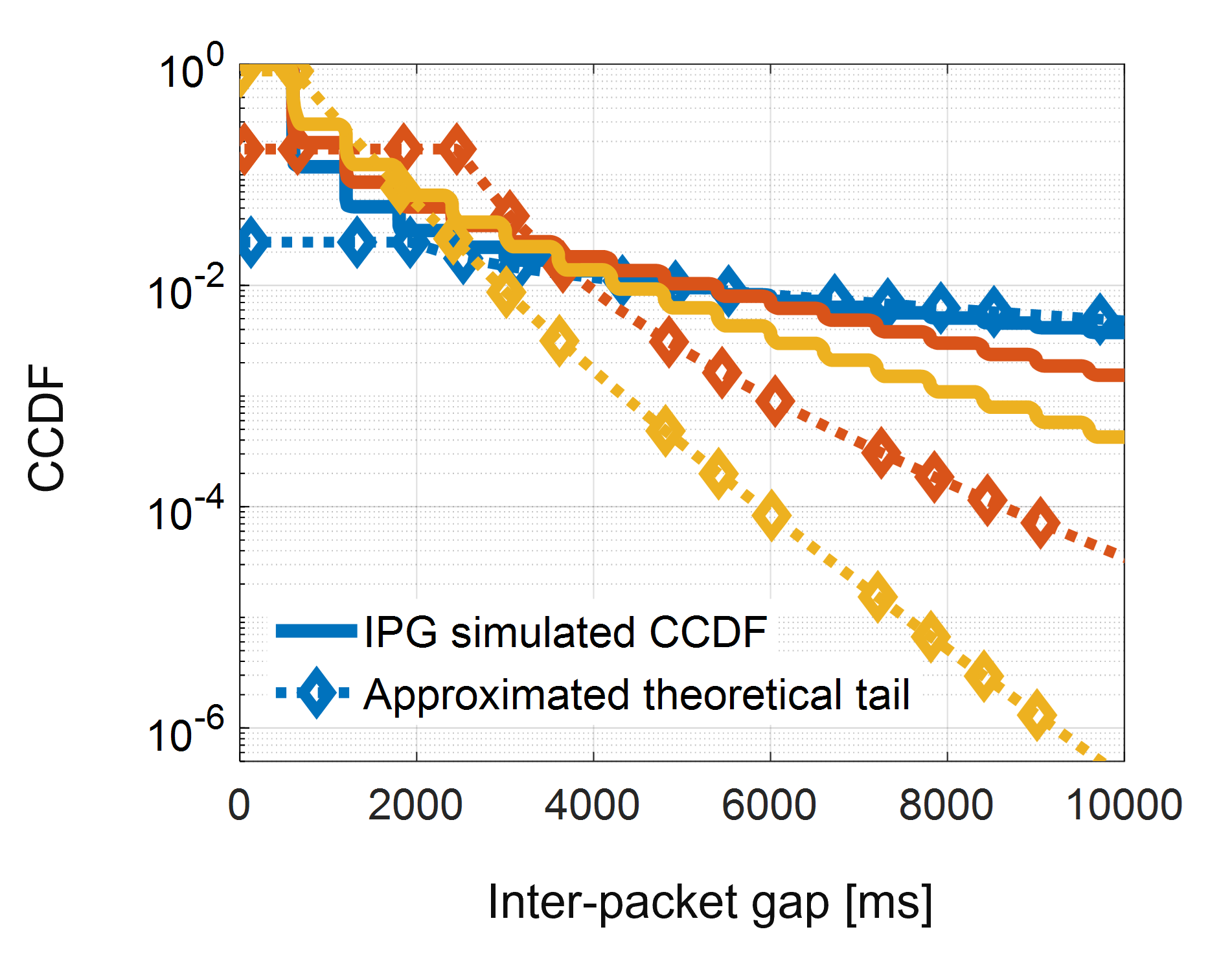}
  \caption{800 VUE/km, 10 MHz using~(\ref{eq_pa})}\label{fig_valid5}
  \vspace{-.15in}
  \end{center}
\end{figure}

For a high vehicle density of 800 VUE/km, using ~(\ref{eq_pa}) leads to a noticeable gap between the approximated and the simulated IPG tail distributions as shown in Fig.~\ref{fig_valid5} for a 10 MHz bandwidth (a similar trend was observed at the 20 MHz). This is because, in highly congested scenarios and with HARQ retransmissions, it is likely that each BSM transmission attempt suffers interference from more than a single dominant interferer. When considering two dominant interferers for each transmission attempt, $P_{\mathrm{a}}$ (for simplicity of presentation, $q_{\mathrm{i}}^{(k)}$ is denoted as $q_{\mathrm{i},k}$ in the rest of this paper) can be updated as:
\begin{equation}\label{eq_updPa}
    P_{\mathrm{a}}=1-\left(1-q_{\mathrm{i},k}^2\right)^{2}.
\end{equation}

\begin{figure}
  \begin{center}
  \includegraphics[width=6.5cm,height=6.5cm,,keepaspectratio]{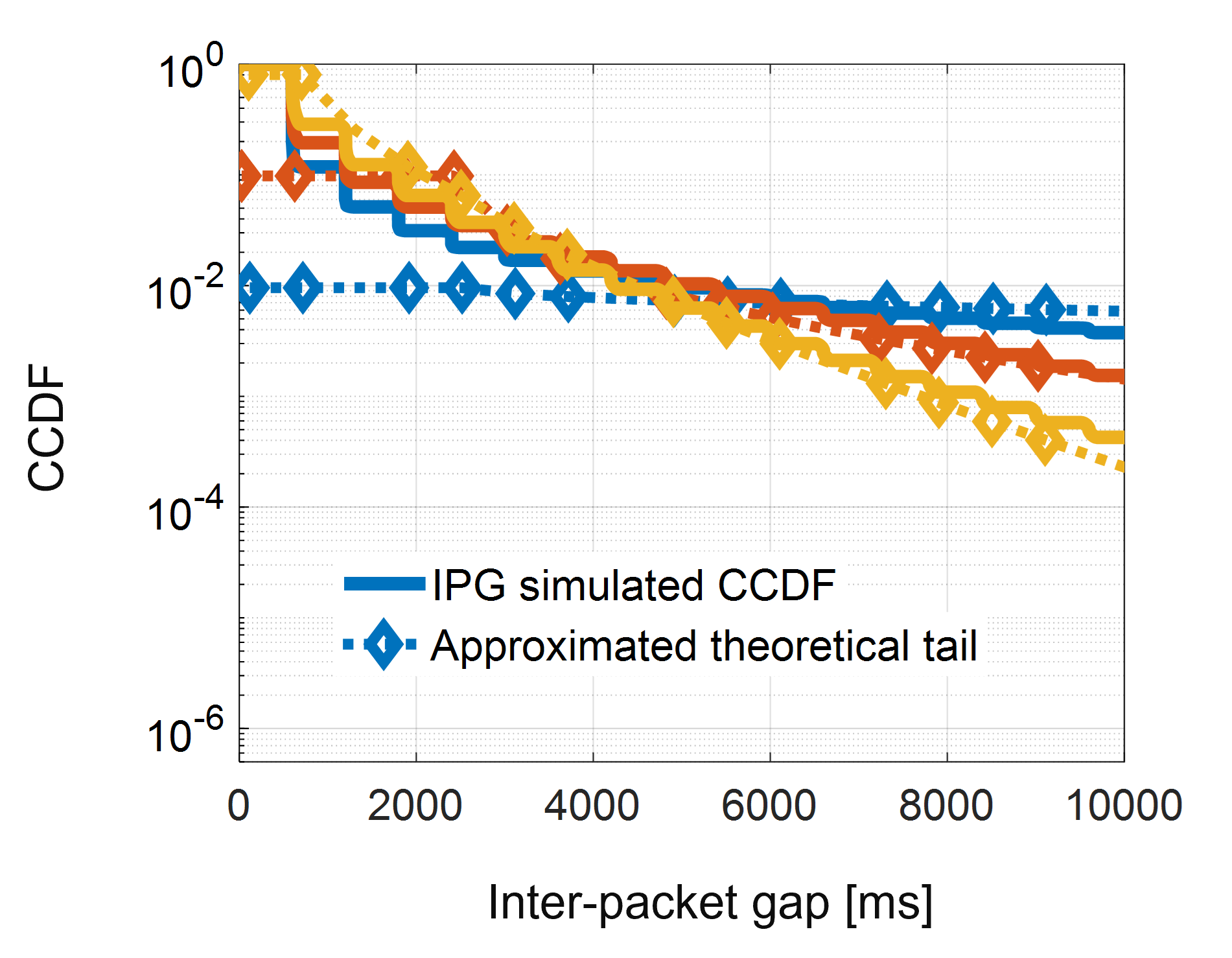}
  \caption{800 VUE/km, 10 MHz.}\label{fig_valid6}
  \vspace{-.15in}
  \end{center}
\end{figure}

\begin{figure}
  \begin{center}
  \includegraphics[width=6.5cm,height=6.5cm,,keepaspectratio]{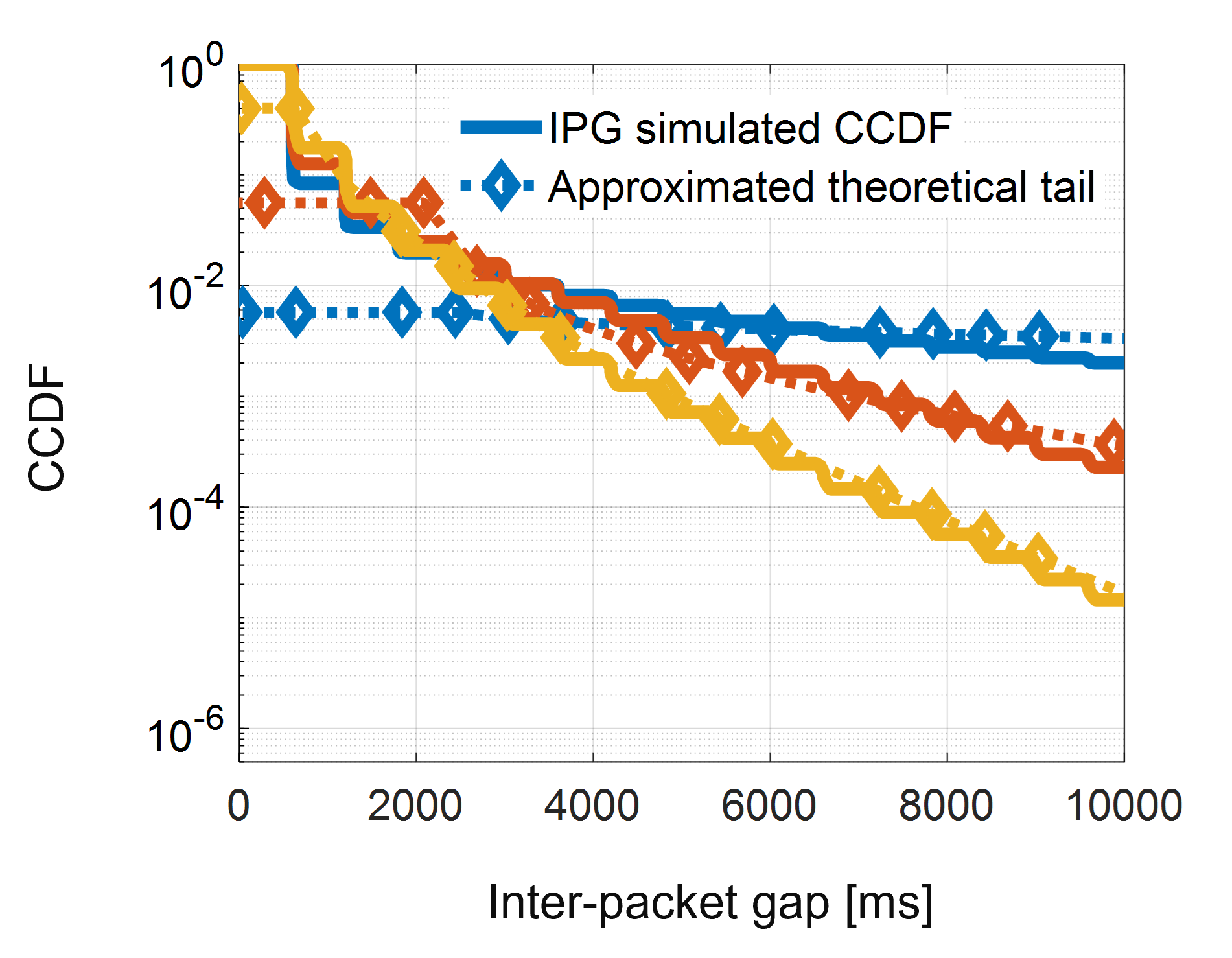}
  \caption{800 VUE/km, 20 MHz.}\label{fig_valid7}
  \vspace{-.10in}
  \end{center}
\end{figure}

The probability of a successful BSM transmission after $k$ transmission opportunities can then be written as:
\begin{equation}\label{eq_updpu}
    \begin{split}
    p_\mathrm{u}^{(k)}=&\left(1-\left(1-q_{\mathrm{i},k}^2\right)^{2}\right)\left(1-q_{\mathrm{i},k}\right)+\\&\left(1-q_{\mathrm{i},k}^2\right)^2q_{\mathrm{i},k}P_{\mathrm{f}}+\left(1-\left(1-q_{\mathrm{i},k}^2\right)^{2}\right)q_{\mathrm{i},k}P_{\mathrm{f}},
    \end{split}
\end{equation}
where the first term represents the event that at least one interferer reselects new VRBs and the transmitter continues using the same VRBs. The second term denotes the event where neither one of the first and second dominant interferers reselects new VRB(s) and the transmitter reselects with probability $q_{\mathrm{i},k}$ and that reselection results in a successful transmission with probability $P_{\mathrm{f}}$. The third term represents the event where at least one of the dominant interferers and the transmitter reselect new VRBs. By substituting (\ref{eq_updpu}) into~(\ref{eq_tk}) the approximated IPG slopes now coincide much better with the simulated IPG tail distributions, as shown in Figs.~\ref{fig_valid6} and~\ref{fig_valid7}.\footnote{For VUE densities that result in a packet generation rate that is not a multiple of 100 ms due to congestion control, then the approximated slope in (\ref{eq_tk}) needs to be updated to account for "slippage events", where a transmitter may miss transmitting in a VRB due to the packet generation rate not being matched to the VRB frequency. This is discussed in \cite{spsjour} without HARQ and a similar approach can be adopted here.} 


\section{Conclusion}\label{sec_conc}
A high-fidelity system-level simulator is used to evaluate the latency of BSM transmission (in terms of IPG) and reliability (in terms of PRR) when HARQ retransmission are used with the SAE-based one-shot method for the sensing-based SPS in C-V2X transmission mode 4. In addition, an analytical model is presented to give a good approximation for the tail distribution of the IPG CCDFs. The study reveals that the impact of HARQ retransmissions on the IPG tail behavior and PRR performance are influenced by several deployment settings, including, but not limited to, bandwidth configuration, vehicle density, and Tx-Rx separation. 

\balance
\bibliographystyle{IEEEtran}
\bibliography{IEEEabrv.bib,Bibliography.bib}
\end{document}